\pdfoutput=1
\documentclass{JINST}
\usepackage{ amssymb }
\usepackage{graphicx}

\providecommand{\color}[2][1]{}
\usepackage{ulem}
\newcommand{\dd}{\mathrm{d}}

\title{Amplitude estimation of a sine function
based on confidence intervals and Bayes' theorem}

\author{Dennis Eversmann$^{a}$,
J\"org Pretz$^{a,b}$\thanks{Corresponding author.}~ and
Marcel Rosenthal$^{a,c}$\\
\llap{$^a$}III. Physikalisches Institut B, RWTH Aachen University, 52056 Aachen, Germany\\
\llap{$^b$}JARA-FAME (Forces and Matter Experiments), Forschungszentrum
J\"ulich und RWTH Aachen University\\
\llap{$^c$}Institut f\"ur Kernphysik, Forschungszentrum J\"ulich, 52425 J\"ulich, Germany\\
E-mail: \email{pretz@physik.rwth-aachen.de}
}

\abstract{
	This paper discusses the amplitude estimation  using data originating from a sine-like function
	as probability density function.
	If a simple least squares fit is used, a significant bias is observed if the amplitude is small
	compared to its error.
	It is shown that a proper treatment using the Feldman-Cousins algorithm of likelihood ratios 
	allows one to construct improved confidence intervals. Using Bayes' theorem a probability density function
	is derived for the amplitude. It is used in an application to show that it leads to better estimates compared to a simple 
	least squares fit.}


\begin{document}
\section{Introduction and motivation}
\begin{figure}[h]
 \includegraphics[width=\textwidth]{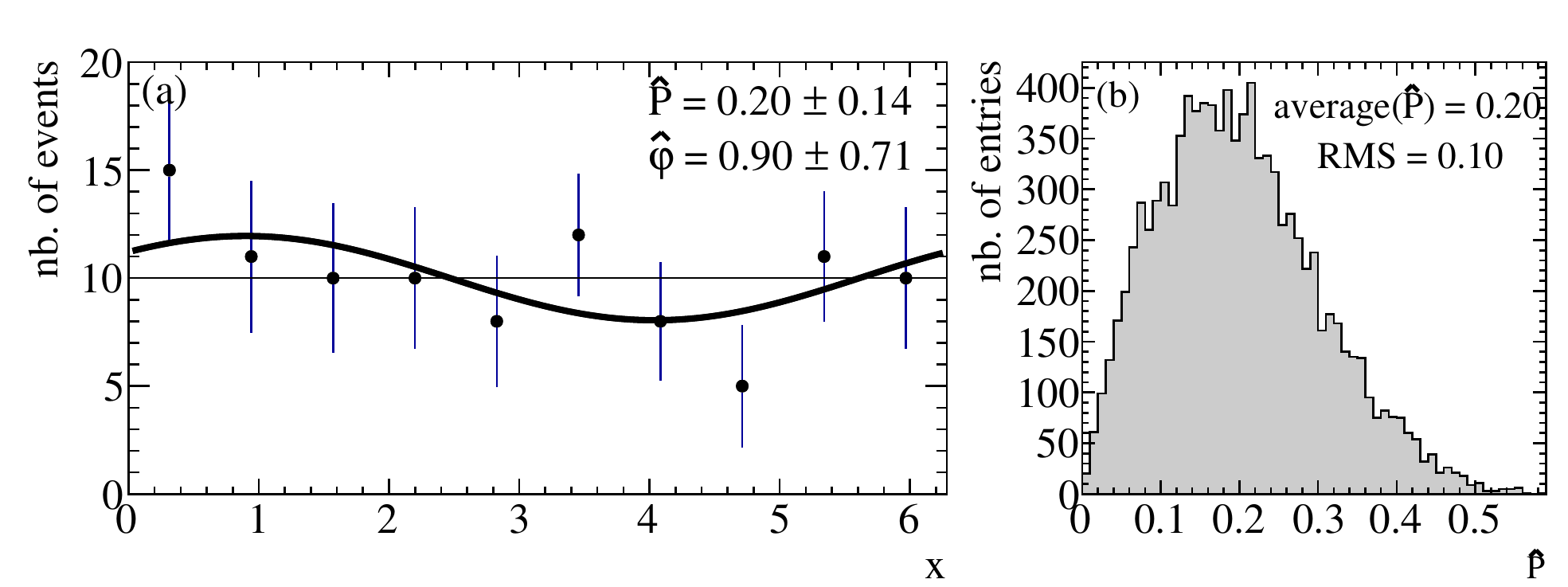}
 \caption{(a): Data points simulated according to eq.~(\protect\ref{f1}) with $N=100$ events
 with $A=0.1$ and $B=0$. The black line shows the result of a least squares fit with 
$\hat P=\sqrt{\hat A^2+\hat B^2} = 0.20\pm 0.14$ and  $\hat \varphi = 0.90\pm 0.71$.
(b): Distribution of $\hat{P}$ for $10000$ simulations. \label{fig:sin01}}
\end{figure}

This paper describes the amplitude estimation of a sine-like function.
In general there is a bias towards an overestimation of the amplitude
which is particularly significant {if the amplitude is small compared to its error}.
Our starting point are data distributed on average according to the following functional form
\begin{equation}\label{f1}
y(x) \propto (1 + A \sin(x) + B \cos(x) ) \, ,
\end{equation}
which can also be written as $(1 + P \cos(x - \varphi))$ with $P = \sqrt{A^2+B^2} \ge 0$ 
and $\varphi = \mbox{atan2}(A,B)$
\footnote{atan2 is the four-quadrant inverse tangent.}.
Distributions like eq.~(\ref{f1}) are widely discussed in signal processing~\cite{alegria,haendel}.
In particle physics they occur in scattering experiments with
polarized beams and/or polarized targets. In this case $y(x)$ is proportional to the counting 
rate depending on the azimuthal angle $x$~\cite{PhysRevC.74.064003}, or,
in case of a precessing polarization vector, $x$ is
proportional to the time~\cite{Eversmann:2015jnk,PhysRevSTAB.17.052803}.
Figure~\ref{fig:sin01} (a) shows a distribution with $N=100$ events. The data 
were randomly generated according to eq.~(\ref{f1}) with $A=0.1$ and $B=0$ and a Poisson statistical 
error was assumed. The black curve shows the result of a least squares fit.

The goal is to determine the amplitude $P=\sqrt{A^2 + B^2}$.
If $P$ is large,  
this is a trivial task 
by just performing a least squares fit. 
For $P$ close to the boundary $P=0$
the task is more difficult.
For small amplitudes  the estimated value $\hat P$ is on average larger than the true $P$.
This is evident, because for $P=0$ the least squares fit in general results in $\hat P>0$. 
In the example of figure~\ref{fig:sin01}~(a) the fit yields $\hat P=0.20 \pm 0.14$. As expected, the statistical error 
is approximately equal to $\sqrt{2/N}$. The exact expression for the error is derived
in appendix~\ref{app:staterr}.
Figure~\ref{fig:sin01}  (b) shows the result of $\hat P$ for 10000 fits to distributions generated with $P=0.1$.
The average $\hat P$ amounts to 0.2 which corresponds to a bias of 0.1.
Interpreting the fit result $\pm 1 \, \sigma$ as a 68\% confidence interval 
for $P$ may lead to coverage in the unphysical region below zero.

\clearpage

The bias of the least squares estimate is discussed and given analytically
in Refs.~\cite{alegria,haendel}.
However, in these references the question of unphysical values for the confidence interval is not addressed.
The main subject of this paper is the application of the Feldman-Cousins algorithm~\cite{Feldman:1997qc} 
to construct proper confidence intervals for $P$ and deriving a probability density function for $P$
making use of Bayes' theorem.
Note that if the phase $\varphi$ is known, the distribution
can be described by a single parameter $P$
%\[ 
% y(x) \propto (1 +  P \cos(x - \varphi)) \, 
%\]
where the  estimated parameter $\hat P$ can be positive or negative although $P\ge 0$.

The paper is organized as follows. Section~\ref{ci} describes the construction of confidence intervals.
In section~\ref{pdf} a probability density function is derived. Section~\ref{app} discusses one application. 

\section{Construction of confidence intervals}\label{ci}
For data distributed according to eq.~(\ref{f1}), the probability distribution function 
for $\hat P=\sqrt{\hat A^2+\hat B^2}\ge0$ can be derived analytically assuming 
that $\hat A$ and $\hat B$ are uncorrelated and normal distributed with means $A$ and $B$,
respectively and variance $\sigma^2$.
The combined distribution for $\hat A$ and $\hat B$ is given by
\begin{eqnarray}
 f(\hat A|A) f(\hat B|B) \dd \hat A \dd \hat B &=& \frac{1}{2 \pi \sigma^2} \, \mathrm{e}^{-(\hat A-A)^2/(2\sigma^2)}
              \mathrm{e}^{-(\hat B-B)^2/(2\sigma^2)} \, \dd \hat A \dd \hat B \, . \label{eq:fAB}
\end{eqnarray}
The transformation to $\hat P=\sqrt{\hat A^2+\hat B^2}$ and $\hat \varphi = \mbox{atan2}(\hat A,\hat B)$
leads, after integration over $\varphi$, to
\begin{equation}\label{fP}
 f(\hat P|P) \dd \hat P = \frac{1}{\sigma^2} \, \mathrm{e}^{-(\hat P^2+P^2)/(2\sigma^2)} 
  \hat P \, I_0\left(\frac{\hat P P}{\sigma^2}\right) \, \dd \hat P
\end{equation}
where $I_0$ is the modified Bessel function of first kind.
Details are given in appendix~\ref{app:rice}.

\begin{figure}
   \includegraphics[width=\textwidth]{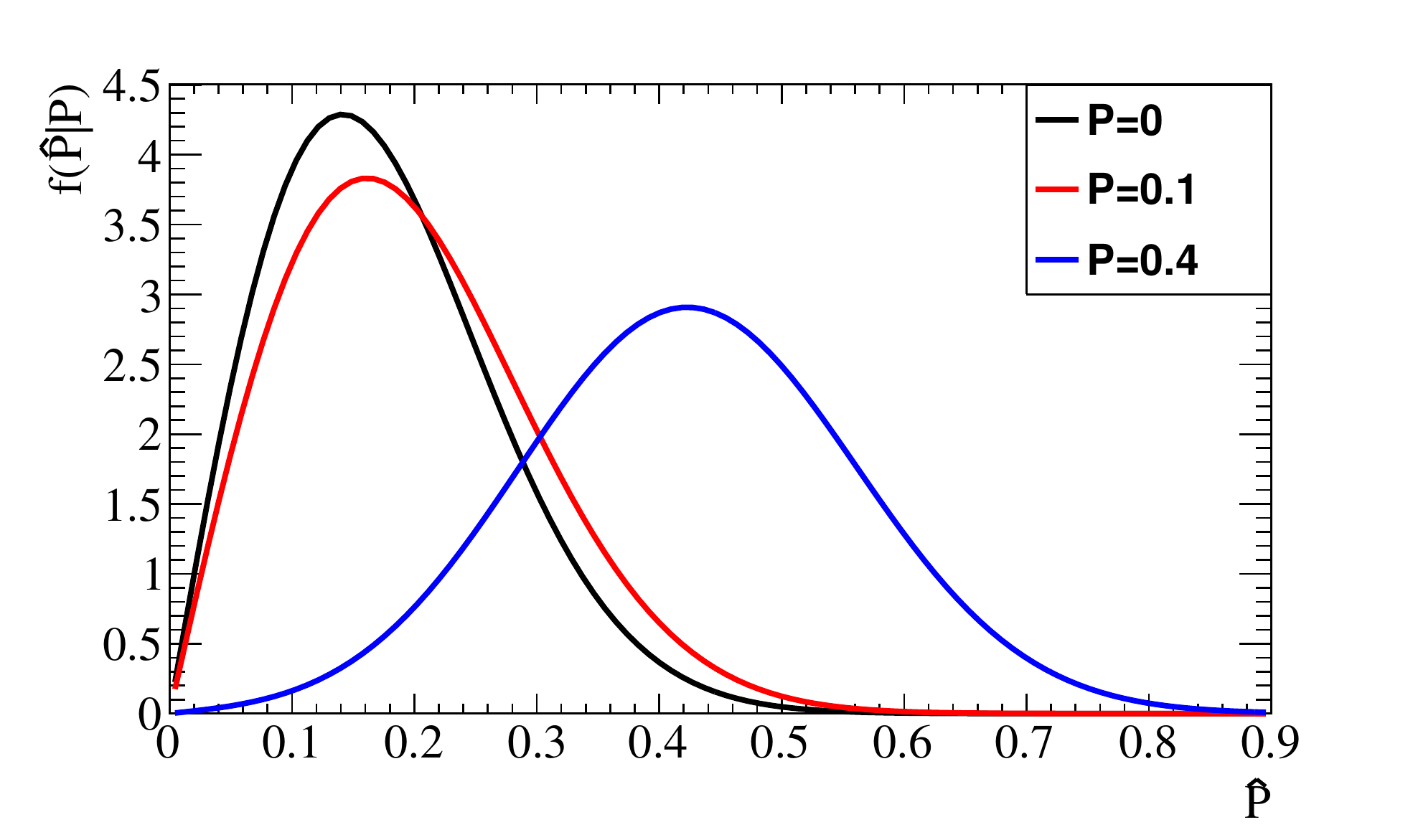}
 \caption{Probability density function $f(\hat P|P)$ for  $P = 0.4, 0.1$ and $0$ for $\sigma = \sqrt{2/100}$.
 \label{fig:proj}}
\end{figure}

Figure~\ref{fig:h2} shows $f(\hat P|P)$ for $\sigma=\sqrt{2/100}$. It is known as the Rice distribution~\cite{2002probability}.
In figure~\ref{fig:proj} $f(\hat P|P)$ is shown for $P = 0.4, 0.1$ and $0$.
For $P=0.4$ the estimated $\hat P$ follows approximately a  normal distribution $N(P,\sigma)$. 
For smaller values of $P$ the bias $(\left< \hat P \right>-P)$ increases. 
In the case of $P=0.1$, one finds $\left< \hat P\right> =0.2$, which agrees 
well with the bias given in Refs.~\cite{haendel,2002probability} and with the
observation in figure~\ref{fig:sin01} (b).
The magnitude of the bias depends not only on $P$ but also on $\sigma$.

\begin{figure}
   \includegraphics[width=\textwidth]{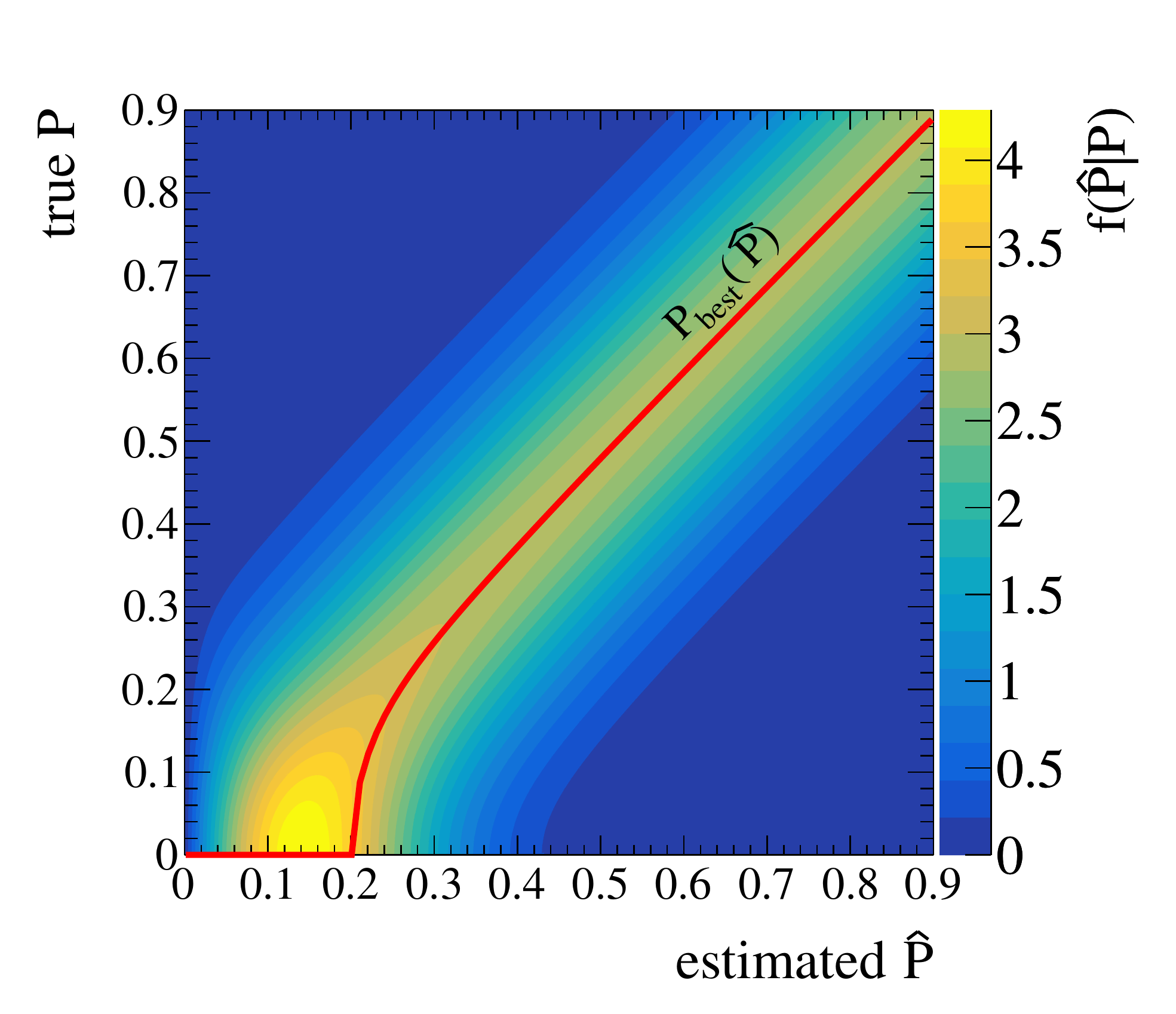}
 \caption{
 Probability density function $f(\hat P|P)$ for $\sigma = \sqrt{2/100}$.
 The red line shows $P_{\mathrm{best}} (\hat P)$.
  \label{fig:h2}}
\end{figure}

In the following the Feldman-Cousins algorithm~\cite{Feldman:1997qc} is used
to construct a confidence interval.
 A general introduction on confidence intervals can be found in
ref.~\cite{james2006statistical}.
More details are also given in appendix~\ref{app:cfi}.
At a given value of $P$ the algorithm selects 
all values of $\hat P$ %in the confidence interval 
for which the ratio
\begin{equation}\label{fR}
  R(\hat P,P) = \frac{f(\hat P|P)}{f(\hat P|P_{\mathrm{best}})}
\end{equation}
has the largest values until the desired coverage of the confidence interval is reached.
$P_{\mathrm{best}}$ denotes the value for which $f(\hat P|P_{\mathrm{best}})$ has its maximum
in the allowed region of $P$, i.e. $f(\hat P|P_{\mathrm{best}}) = \max\{f(\hat P|P)\}$.
 $P_{\mathrm{best}}$ as a function of $\hat P$ is shown in figure~\ref{fig:h2}
as a red line.

\begin{figure}[h]
 \includegraphics[width=0.48\textwidth]{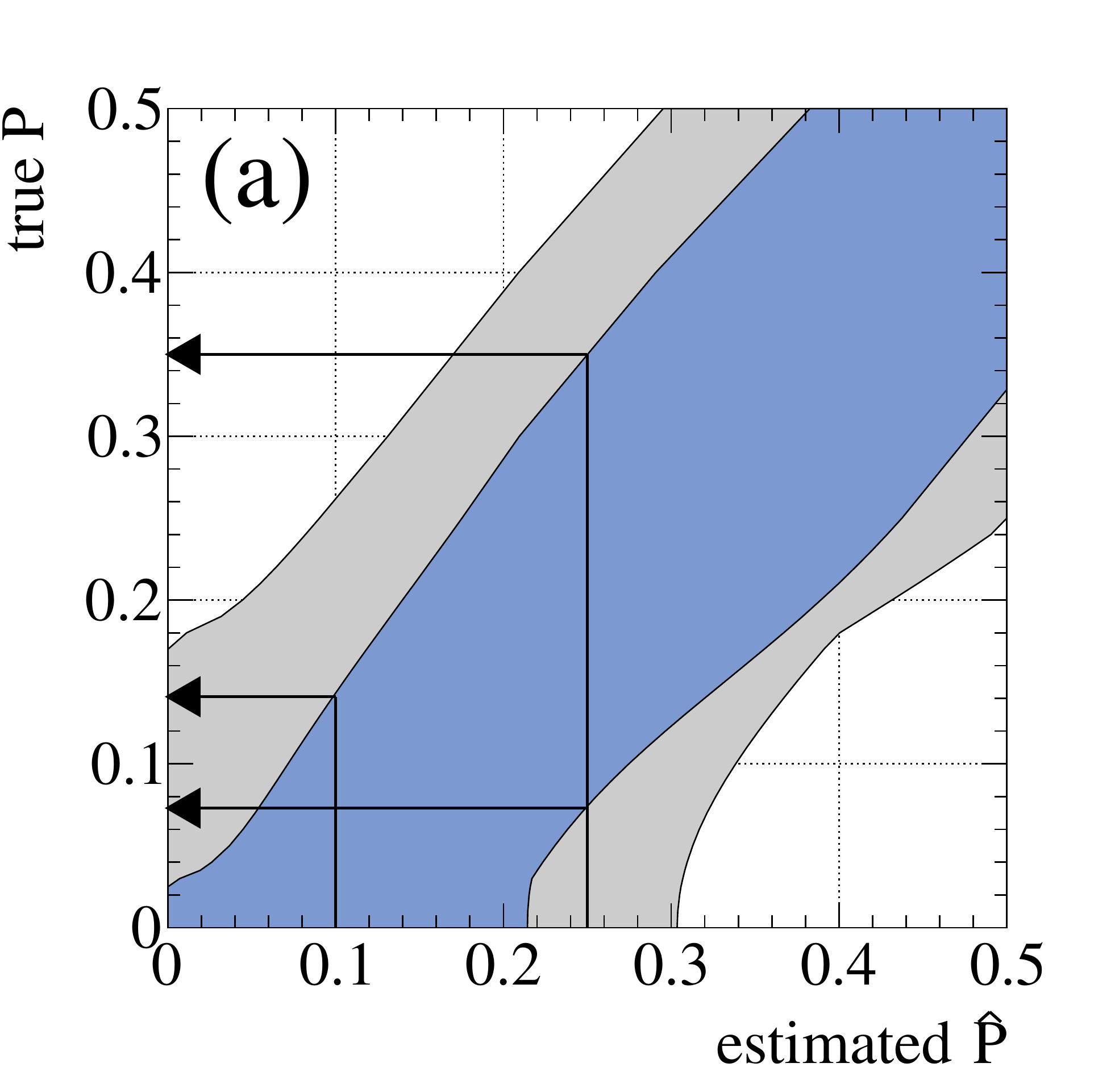}
 \includegraphics[width=0.48\textwidth]{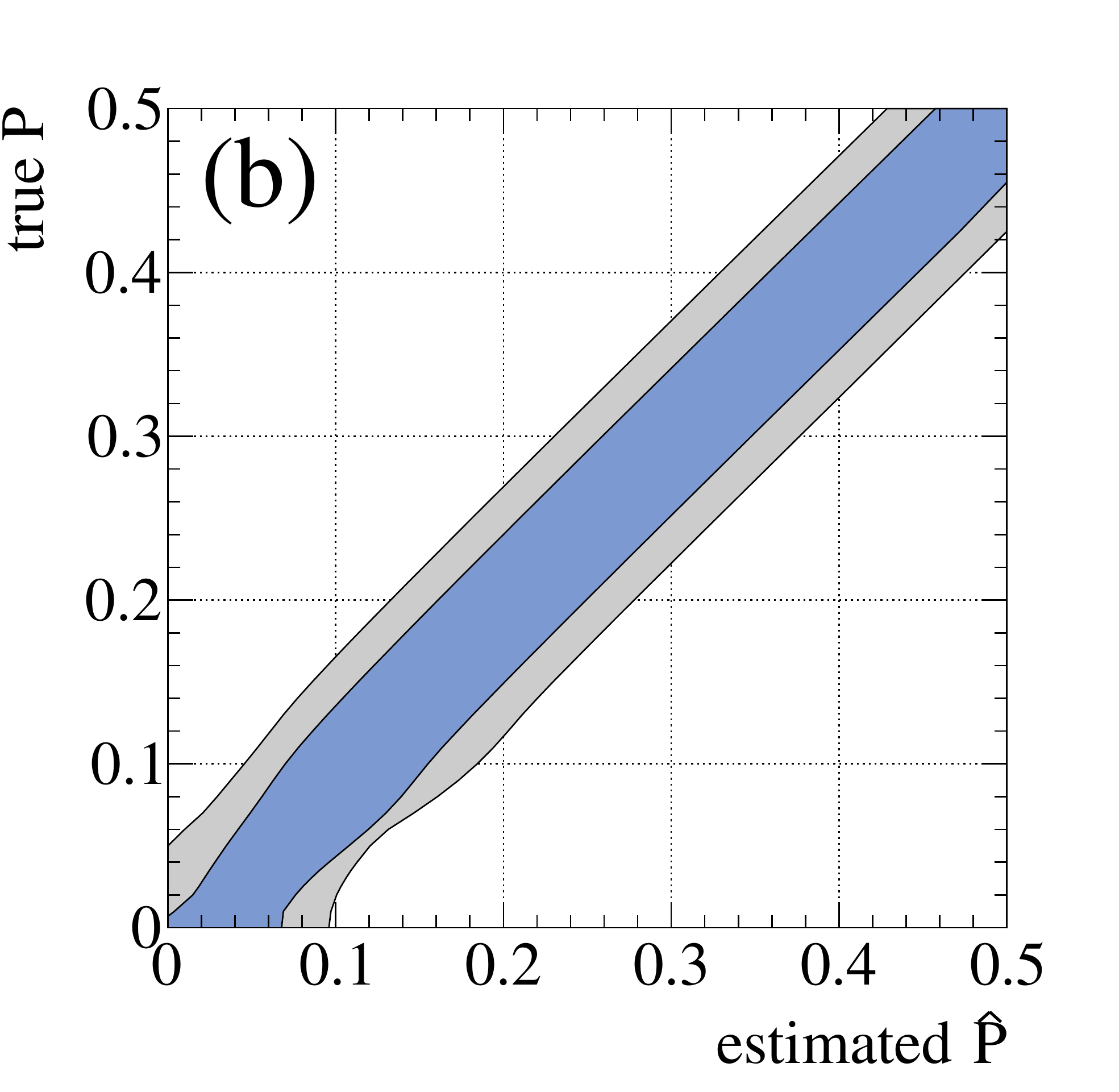}
\caption{Confidence intervals for 68\% (blue) and 90\% (gray) coverage
for $N=100$ events (a) and $N=1000$ events (b). 
In figure (a) for a measured $\hat P=0.1 (0.25)$ the 68\% confidence
interval for $P$ is $[0,0.14] \,([0.073,0.35])$
as indicated by the arrows.
\label{fig:cf}}
 \end{figure}

Figure~\ref{fig:cf} shows the 68\% (blue) and 90\% (gray) confidence intervals for $N=100$ (a)
and $N=1000$ events (b).
In the case of $N=100$ and a measured value $\hat P=0.1$ the 68\%
confidence interval for $P$ ranges from 0 to 0.14.
At larger values of $\hat P$ the 68\% confidence interval coincides with 
the Gaussian error expectation $\left[P-\sqrt{2/N},P+\sqrt{2/N}\right]$. 
For larger $N$, the transition to a Gaussian confidence interval occurs at smaller values of $P$.
By construction, the confidence intervals only contain values in the physical region $P\ge0$.

\section{Derivation of probability density function}\label{pdf}
The previous section described how a confidence interval can be defined
for $P$ given a measured $\hat P$. If the amplitude $P$ is the final result of the experiment, this is sufficient.
However, in many applications $P$ is used as an input in a subsequent analysis. In this case, 
it is desirable to have a probability density function (pdf) for $P$. Unfortunately, it is not
directly possible to construct such a pdf without further assumptions.
To proceed, we make use of the Bayes' theorem with a constant prior probability for $P$.
This leads to the following pdf:
\[
\tilde f(P|\hat P) = \frac{f(\hat P|P)}{\int_0^\infty f(\hat P|P) \dd P }\, .
\]
Note that $\tilde f(P|\hat P) \propto R(\hat P,P)$ for a fixed $\hat P$.
In case of $\hat P=0.1$ and $\sigma=\sqrt{2/100}$ the interval $[0,0.16]$ covers 68\%, i.e. $\int_0^{0.16} \tilde f(P|\hat P) \dd P= 0.68$. 
For comparison, the corresponding
Feldman-Cousins interval is $[0,0.14]$.
In the next section an application is discussed where we will make use of this pdf.

 \section{Application}\label{app}
\begin{figure}
 \includegraphics[width=\textwidth]{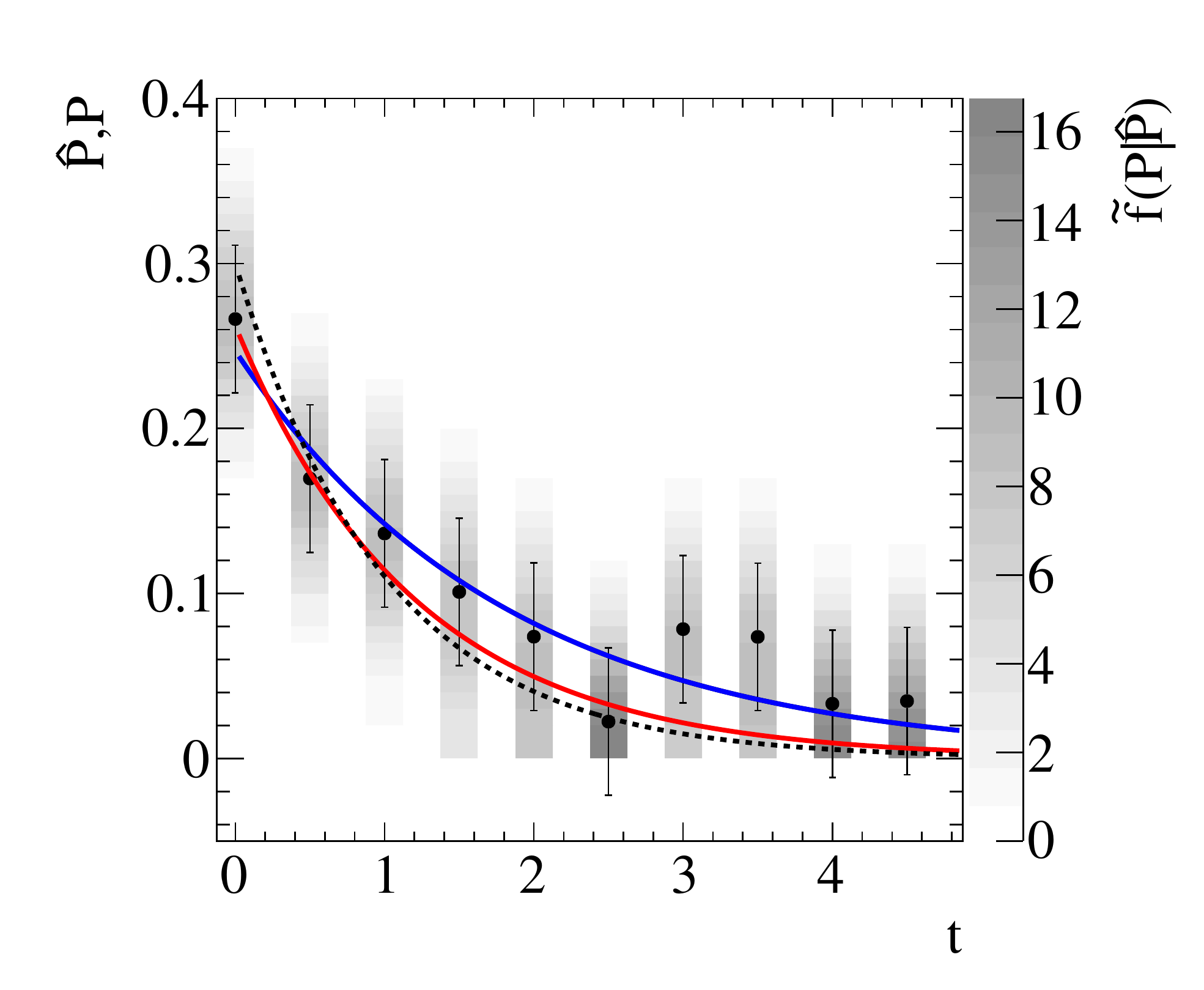}
  \caption{
 Dotted line: $P(t) = C \mbox{e}^{-t/\tau}$ with $\tau =1$ and $C=0.3$.
 Data points: random values according to $f(\hat P|P(t_i)), \quad i=1,\dots,
 10$  with $\sigma =\sqrt{2/1000}$.
 Blue line: result of a least squares fit to the data points.
 Vertical bands: probability distribution $\tilde f(P|\hat P)$ for true $P$ for the given generated $\hat P$.
 Red line: Result of a likelihood fit to these probability distributions.
 \label{fig:sin_expdecay}}
 \end{figure}

\begin{figure}
 \includegraphics[width=\textwidth]{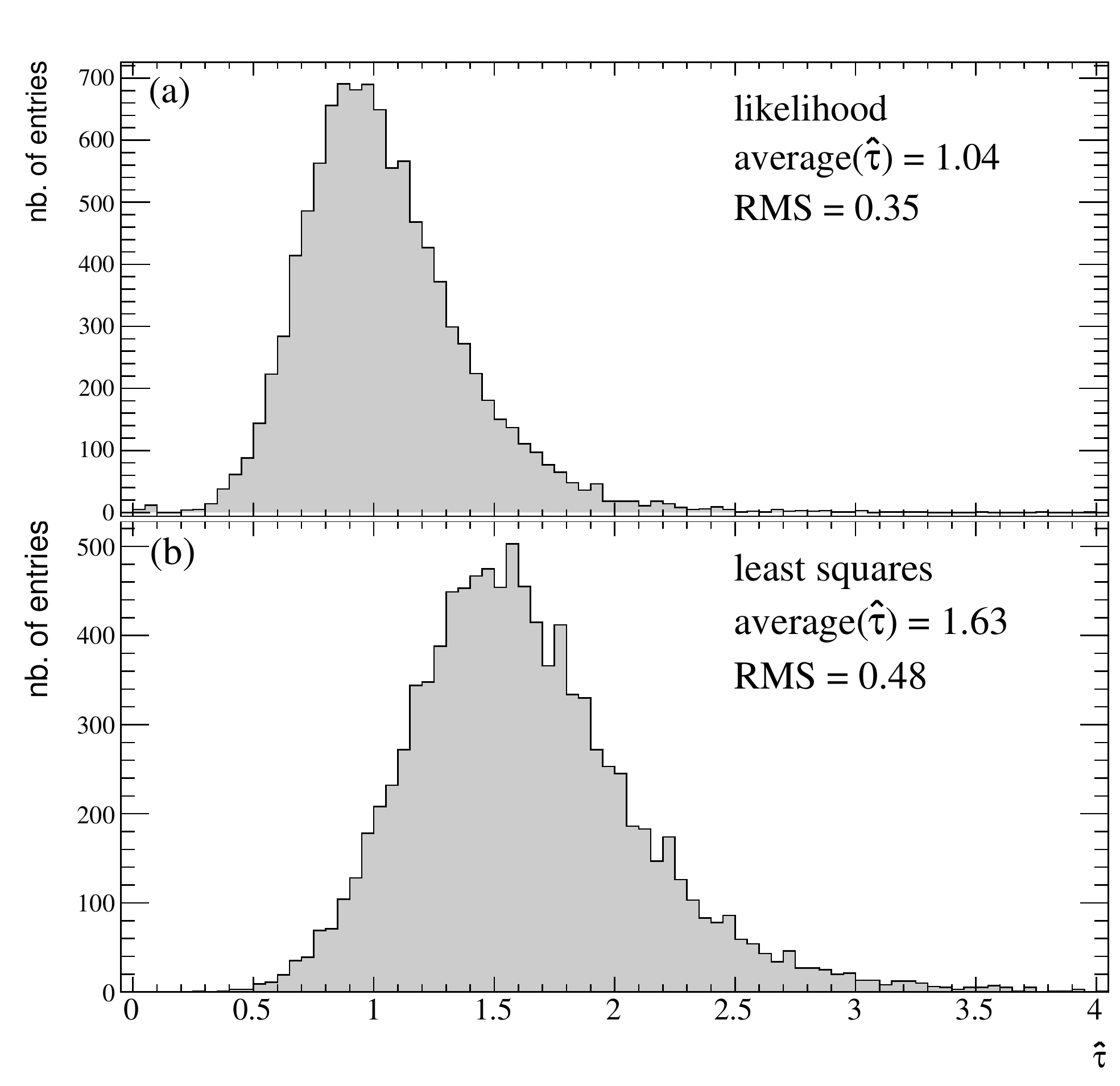}
  \caption{
 Distribution of $\hat \tau$ for 10000 simulations.
Results from likelihood fit using the probability distributions (a),
 and a least squares fit (b).
 \label{fig:tau_dist}}
 \end{figure}

 We study the case where the amplitude decays exponentially with time.
 In figure~\ref{fig:sin_expdecay} the dotted line shows an exponential function
 $C \mbox{e}^{-t/\tau}$ with a decay constant $\tau =1$ and amplitude $C=0.3$. 
 Data were generated at ten different times $t$ following the distribution $f(\hat P|P)$ in eq.~(\ref{fP})
 with $P = C \mbox{e}^{-t/\tau}$, $\sigma=\sqrt{2/N}$ and $N=1000$.
 These values are displayed as data points.
 The blue curve shows the result of a least squares fit to these data points.
The vertical bands at each $t$-bin show the pdf  
  $\tilde f(P|\hat P)$ as a function of $P$, where $\hat P$ is the generated value (i.e. the data
 point). 
The red curve shows the result of a likelihood fit with the likelihood function
 \[
  \mathcal{L} = \Pi_{i=1}^{N_{bin}} \, \, \tilde f(C \mbox{e}^{-t_i/\tau}|P_i)
 \]
varying $\tau$ and $C$ to maximize $\mathcal{L}$.
In this example the likelihood fit yields 1.20$\pm$0.37 (red line) for $\hat \tau$ compared to 1.81 $\pm$ 0.51 (blue line) 
for the least squares fit to the black points.
Figure~\ref{fig:tau_dist} shows the result of 10000 simulations.
The average of $\hat \tau$  is 1.04 for the likelihood and 1.63 for the least squares fit.
On average the likelihood result has a bias of 0.04/0.35=0.11 of its
 statistical error, whereas the bias for the least square fit is
 0.63/0.48=1.3.
This  proves  that the likelihood fit using $\tilde f(\hat P|P)$ as a pdf
gives a result closer to the true value $\tau =1$. 
Since the likelihood method is only asymptotically unbiased ($N_{\mathrm{bin}}\rightarrow \infty$),  
the  small bias  even decreases if the number of $t$-bins is increased.

\section{Conclusions}
Parameter estimations play an important role in all
area of science. 
Estimates of theses parameters obtained from least squares fits
assuming Gaussian errors are often biased even for simple scenarios
like the estimation of an amplitude of a sine-function.
This may in addition introduce coverage in non-physical regions of the parameter.

In this paper we made use of the Feldman-Cousins algorithm to construct confidence intervals 
for the amplitude $P$ of a sine-function
covering only the allowed region $P\ge 0$. 
Further, a probability density function (pdf) for the amplitude was derived applying the Bayes' theorem.
In an application it was shown that using this pdf
leads to better fit results compared to a simple least squares fit.

\acknowledgments
We would like to thank Colin Wilkin for the careful reading of the manuscript % and helpful comments
and members of JEDI collaboration for stimulating discussions on the subject.

\newpage
\appendix

\section{Statistical error of amplitude  and phase}\label{app:staterr}
Starting from the probability density function
\[
  y(x) = (1+ P \cos(x-\varphi) ) \, ,\quad 0 \le x < 2\pi \, ,
\]
the log-likelihood function reads
\[
  \ell = \sum_{i=1}^{N} \log(1+ P \cos(x_i - \varphi) ) \, .
\]
Using the second derivatives
\begin{eqnarray}
  \frac{\partial^2 \ell}{\partial P^2} &=& - \sum_i \frac{\cos^2 (x-\varphi)}{(1+P \cos (x-\varphi))^2}    \, ,\\
  \frac{\partial^2 \ell}{\partial \varphi^2} &=& - \sum_i \frac{-P^2 \sin^2(x-\varphi)}{(1+P \cos (x-\varphi) )^2}    \, ,\\
  \frac{\partial^2 \ell}{\partial P \partial \varphi} &=& - \sum_i \frac{ \sin(x-\varphi) }{(1+P \cos (x-\varphi) )^2}    \, 
\end{eqnarray}
and their expectation values
\begin{eqnarray*}
  \left< \frac{\partial^2 \ell}{\partial P^2}\right> &=& N \, \int_{0}^{2\pi} \frac{\partial^2 \ell}{\partial P^2}(x) \, y(x) \dd x \\
               &=& N \, \frac{1 - \sqrt{1/(1-P^2)}}{P^2}  \stackrel{P^2 \ll 1}{\approx}  -\frac{N}{2} \left( 1 + \frac{3}{4} P^2 \right) \, ,\\
   \left<\frac{\partial^2 \ell}{\partial \varphi^2}  \right>&=& N \, \frac{1 - \sqrt{1/(1-P^2)}}{\sqrt{1/(1-P^2)}}  \stackrel{P^2 \ll 1}{\approx}  -\frac{N P^2}{2}  \, ,\\
   \left<\frac{\partial^2 \ell}{\partial P \partial \varphi}  \right> &=& 0   \, .
\end{eqnarray*}
the errors on  $P$ and $\varphi$ can be calculated.
To $\mathcal{O}(P^2)$ they are given by
\begin{eqnarray*}
  \sigma_P^{-2} &=& \frac{N}{2} \left( 1 + \frac{3}{4} P^2 \right) \, , \\
  \sigma_\varphi^{-2} &=& \frac{N P^2}{2}  \, . 
\end{eqnarray*}
Note that for a sufficient large number of bins the errors derived here
for the unbinned likelihood method coincide with the errors 
of the least squares fit.

\section{Derivation of the Rice distribution}\label{app:rice}
Starting from eq.~(\ref{eq:fAB})
\begin{eqnarray}
 f(\hat A|B) f(\hat B|B) \dd \hat A \dd \hat B &=& \frac{1}{2 \pi \sigma^2} \, \mathrm{e}^{-(\hat A -A )^2/(2\sigma^2)}
              \mathrm{e}^{-(\hat B  -B)^2/(2\sigma^2)} \, \dd \hat A \, \dd \hat B \, ,
\end{eqnarray}
the transformation to $\hat P$ and $\hat \varphi$ yields
\begin{eqnarray}
 f(\hat P,\hat \varphi|P,\varphi) \, \dd \hat P \, \dd \hat \varphi &=& 
\frac{1}{2 \pi \sigma^2} \, \mathrm{e}^{-(\hat P \sin \hat \varphi -P \sin \varphi)^2/(2\sigma^2)}
              \mathrm{e}^{-(\hat P \cos \hat \varphi - P \cos \varphi)^2/(2\sigma^2)} \, \hat P \dd \hat P \dd \hat \varphi \, \\
&&\frac{1}{2 \pi \sigma^2} \, \mathrm{e}^{-(\hat P^2  +P^2)/(2\sigma^2)}
              \mathrm{e}^{-(2 \hat P P (\sin\varphi \sin \hat \varphi + \sin\varphi \sin \hat \varphi)/(2\sigma^2)} 
              \, \hat P \dd \, \hat P \, \dd \hat \varphi \, 
\end{eqnarray}

Using
\begin{equation}
\int_0^{2 \pi} \mbox{e}^{(2 P \hat P (\sin(\varphi)\sin(\hat \varphi) + \cos(\varphi) \cos(\hat \varphi))/(2/\sigma^2)} \dd \hat \varphi 
    = 2 \pi I_0\left(\frac{P \hat P}{\sigma^2} \right)
\end{equation}
the integration over $\hat \varphi$ results in
\begin{equation}
   f(\hat P|P) = \frac{1}{\sigma^2} \, \mathrm{e}^{-(\hat P^2  +P^2)/(2\sigma^2)} \, I_0\left(\frac{P \hat P}{\sigma^2} \right) \, \hat P \, .
\end{equation}
$I_0$ is the modified Bessel function of first kind.

\section{Details of the Feldman-Cousins confidence interval construction}\label{app:cfi}
\begin{figure}[hp]
\includegraphics[width=\textwidth]{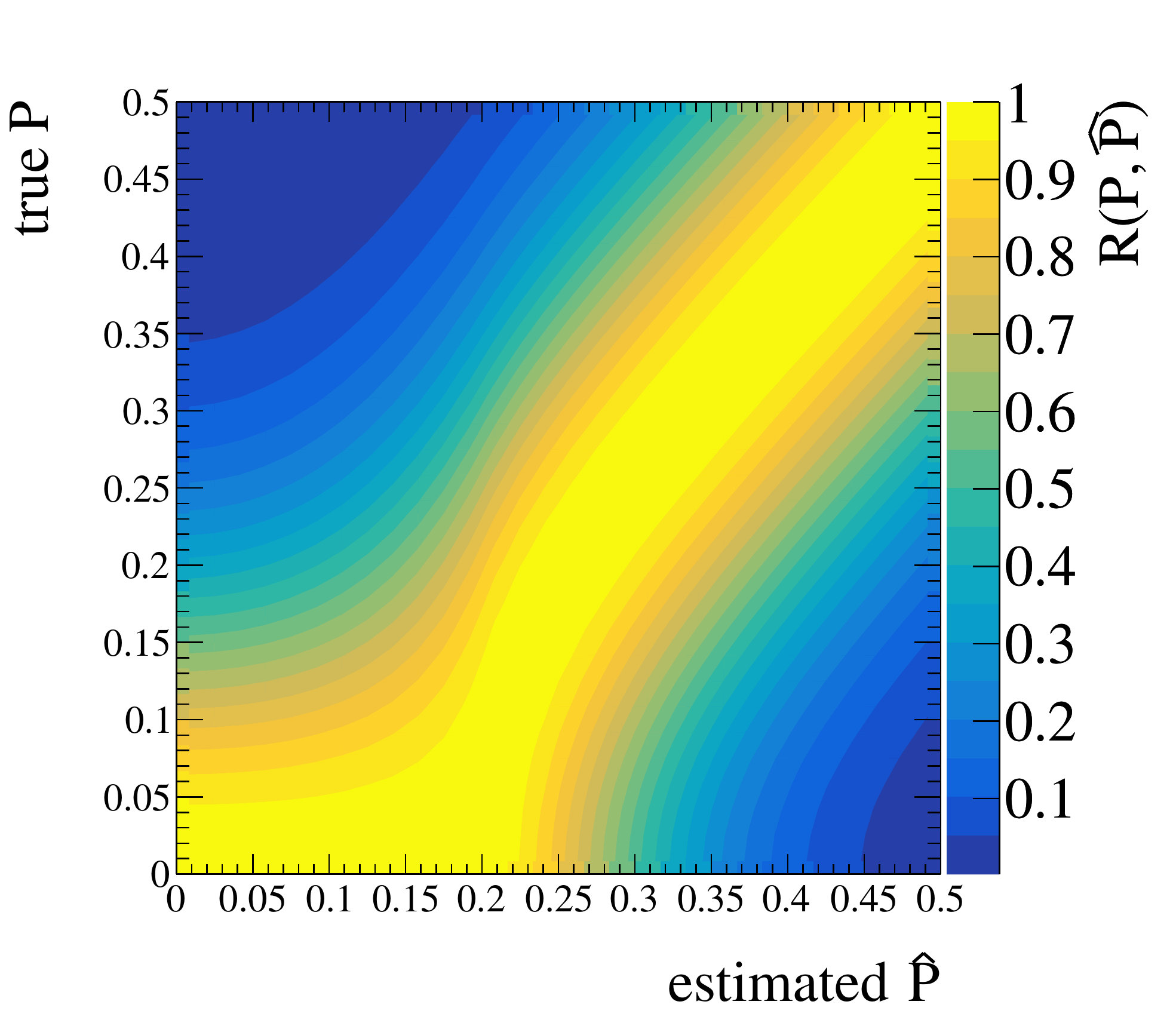}
 \caption{
The likelihood ratio $R(\hat P,P)$ for $\sigma=\sqrt{2/100}$.
\label{fig:fR}}
\end{figure}

\begin{figure}[hp]
 \includegraphics[width=\textwidth]{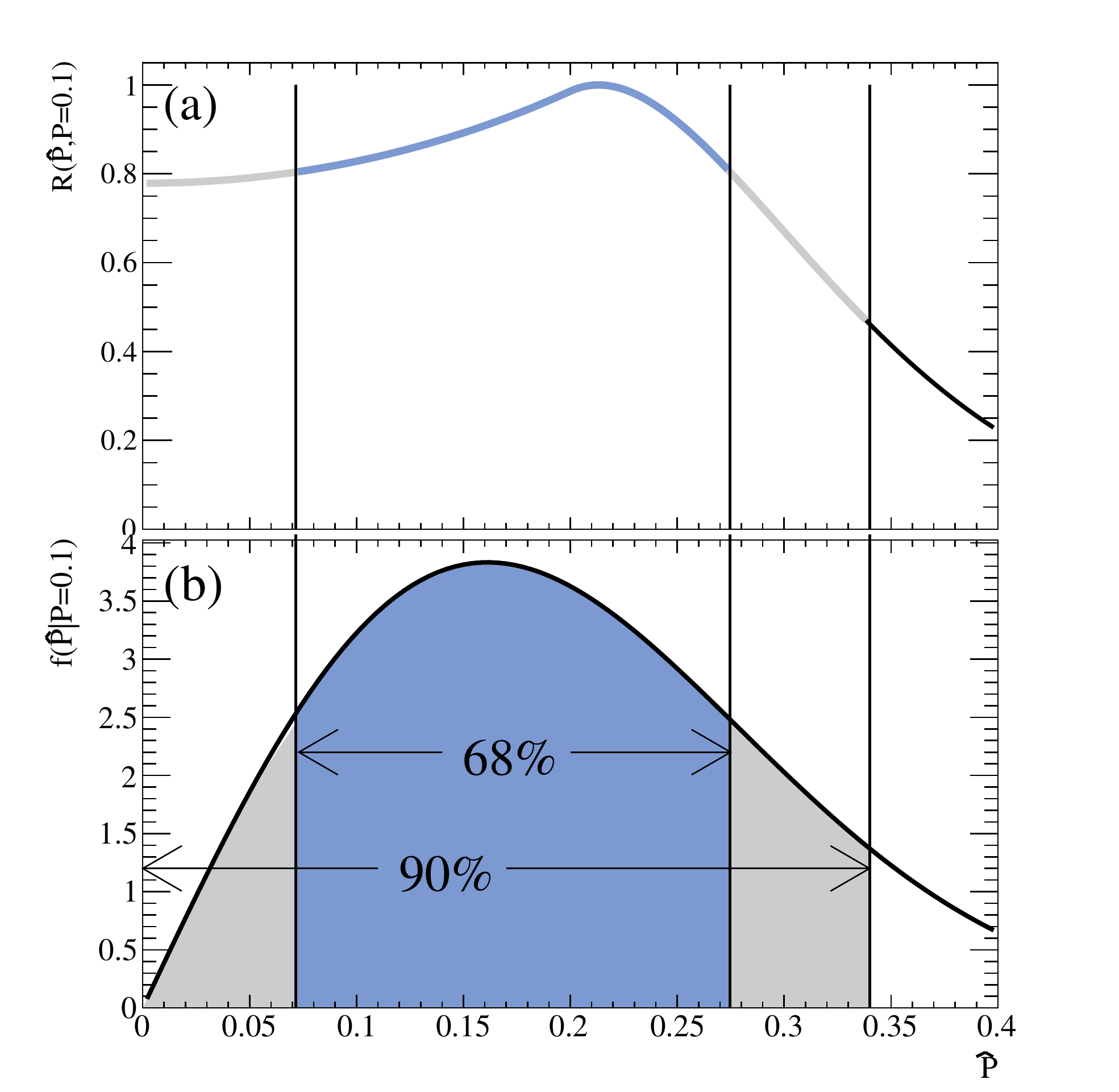}
\caption{Construction of confidence interval. 
(a): The likelihood ratio $R(\hat P,P=0.1)$ (see eq.~(\protect\ref{fR})).
(b):The  probability density function $f(\hat P|P=0.1)$ 
 for $\sigma=\sqrt{2/100}$.
 Starting from the largest value of $R$ in the upper plot
 all values of $\hat P$ are included until the desired confidence interval is reached.
 \label{construction_fc}}
\end{figure}

This appendix shows how confidence intervals are constructed in practice.
The starting point is the likelihood ratio $R(\hat P,P)$ (eq.~(\ref{fR})) shown in 
figure~\ref{fig:fR}.
For each value of true $P$ a lower and upper limit of $\hat P$ is obtained
in the following way.
For a given $P$ one starts at the largest value of $R$ (see
figure~\ref{construction_fc}~(a) for $P=0.1$).
All values of $\hat P$ are included in the interval until 
the desired coverage is reached (see
figure~\ref{construction_fc}~(b)).
In the example given in figure~\ref{construction_fc} the 68\% interval $[\hat P_{\mathrm{min}},\hat P_{\mathrm{max}}]$
is $[0.07,0.27]$
and the 90\% interval is $[0,0.34]$.
 Repeating this procedure for all values of $P$ yields the boundary lines of the colored areas in figure~\ref{fig:cf}. 
This construction makes no use of the measured value.

Given a measurement $\hat P$ 
the confidence interval for $P$ is given by the two
values on the $y$-axis where the vertical line starting at $\hat P$
intersects the boundary lines.
This is indicated in Fig.~\ref{fig:cf}~(a)
for $\hat P=0.1$ and $\hat P=0.25$ for the 68\% intervals.

\clearpage

\bibliography{/home/pretz/bibtex/literature_edm.bib,/home/pretz/bibtex/statistics.bib}

\bibliographystyle{ieeetr}

\end{document}